\documentclass[twocolumn,showpacs,preprintnumbers,amsmath,amssymb]{revtex4}
\usepackage{graphicx}% Include figure files
\begin{document}

\title{Four-point resistance of individual single-wall carbon nanotubes}

\author{B. Gao,$^1$ Y.F. Chen,$^2$ M.S. Fuhrer,$^2$ D.C. Glattli,$^{1,3}$ A. Bachtold$^{1*}$}

\address{
$^1$ Laboratoire Pierre Aigrain, Ecole Normale Sup\'{e}rieure, 24
rue Lhomond, 75231 Paris 05, France. $^2$ Department of Physics
and Center for Superconductivity Research, University of Maryland,
College Park, Maryland 20742, USA. $^3$ SPEC, CEA Saclay, F-91191
Gif-sur-Yvette, France.}

\begin{abstract}
We have studied the resistance of single-wall carbon nanotubes
measured in a four-point configuration with noninvasive voltage
electrodes. The voltage drop is detected using multiwalled carbon
nanotubes while the current is injected through nanofabricated Au
electrodes. The resistance at room temperature is shown to be
linear with the length as expected for a classical resistor. This
changes at cryogenic temperature; the four-point resistance then
depends on the resistance at the Au-tube interfaces and can even
become negative due to quantum-interference effects.

\end{abstract}

\vspace{.3cm} 

\date{ \today}
\maketitle

Transport measurements are a powerful technique to investigate
electronic properties of molecular systems \cite{Joachim}. Most
often, individual molecular systems are electrically attached to
two nanofabricated electrodes. However, such two-point experiments
do not allow the determination of the intrinsic resistance that
results from scattering processes involving e.g. phonons or
disorder. Indeed, the resistance is mainly dominated by poorly
defined contacts that lie in series. A solution to eliminate the
contribution of contacts has been found with scanning probe
microscopy techniques
\cite{Bachtold,Pablo,Freitag,Gudiksen,Yaish}, which enable the
measurement of resistance variations along long systems such as
nanotubes, but these techniques have only been applied at room
temperature. The standard method to determine the intrinsic
resistivity of macroscopic systems is the four-point measurement.
The application of this technique to molecular systems is
challenging however, since the electrodes used so far have been
invasive. For example, nanofabricated electrodes were shown to
divide nanotubes into multiple quantum dots \cite{Bezryadin}.

We report four-point resistance measurements on single-wall carbon
nanotubes (SWNT) using multiwalled carbon nanotubes (MWNT) as
noninvasive voltage electrodes (Fig. 1(a)). We find that SWNTs are
remarkably good one-dimensional conductors with resistances as low
as 1.5 k$\Omega$ for a 95 nm long section. The nanotube resistance
is shown to linearly increase with length at room temperature, in
agreement with Ohm's law. At low temperature, however, the
resistance can become negative and the amplitude then depends on
the resistance at the Au-SWNT interfaces. In this regime,
four-point measurements can be described by the
Landauer-B\"{u}ttiker formalism taking into account
quantum-interference effects.

\begin{figure}
\includegraphics{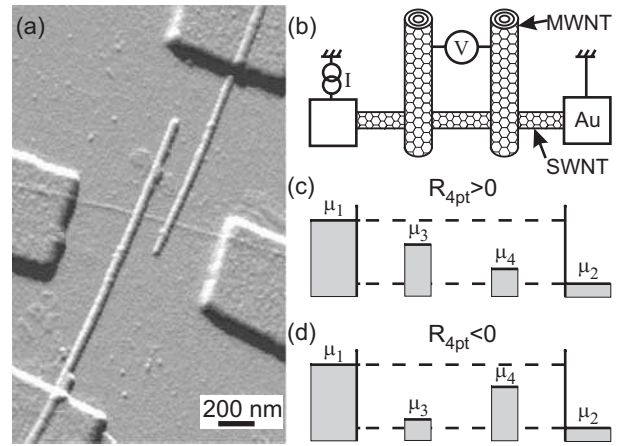}
 \caption{
(a) Atomic force microscopy image of a SWNT contacted by 2 MWNTs
and 2 Au electrodes. (b) Schematic of the $R_{4pt}$ measurement.
$R_{4pt}$ is measured in the linear regime with $eIR_{2pt}$ below
$kT$. (c,d) Levels of the electrochemical potential for the $4$
electrodes that give a positive $R_{4pt}$ in (c) and a negative
$R_{4pt}$ in (d).
 }
\end{figure}

Before discussing these measurements, we briefly review the basic
physics of four-point measurements (Fig. 1(b)). In the diffusive
incoherent limit, the four-point resistance $R_{4pt}$ of SWNTs,
characterized by 4 conducting channels, is given by \cite{Remark1}
%====================================================================
\begin{equation}
R_{4pt}=\frac{h}{4e^2}\frac{L}{l_e}
\end{equation}
%====================================================================
with $l_e$ the elastic mean-free path and $L$ the separation
between the voltage electrodes. Equation 1 describes the intrinsic
resistance generated by the electronic backscattering along the
nanotube.

We next remind how $R_{4pt}$ can deviate from equation 1. The
transmission to the voltage electrodes should be made weak to
suppress an additional resistance contribution that results from
electrons entering the voltage electrodes. Such electrons are
replaced by electrons that scatter into either direction of the
tube, enhancing the resistance \cite{Remark}.

By introducing quantum-interference effects, four-point
measurements can give striking results such as negative $R_{4pt}$.
This is best seen using the B\"{u}ttiker formula
\cite{Buttiker1,Buttiker2,Datta}, which is convenient for the
description of multi-terminal conductors. The current $I_{\alpha}$
in each electrode is related to the electrochemical potential
$\mu_{\beta}$ of other electrodes by
%====================================================================
\begin{equation}
I_{\alpha}=\frac{4e^2}{h}\sum_{\beta}T_{\beta\alpha}\mu_{\alpha}-T_{\alpha\beta}\mu_{\beta}
\end{equation}
%====================================================================
with $T_{\alpha\beta}$ the total transmission between the $\alpha$
and the $\beta$ electrodes (Fig. 1(c,d)). The condition \mbox{$
I_{3}=0$} for a voltage probe gives \mbox{
$\mu_{3}=(T_{31}\mu_{1}+T_{32}\mu_{2})/(T_{31}+T_{32})$}. The
transmission between electrodes 3 and 4 has been neglected since
it corresponds to a second-order process. The potential of the
voltage electrode $\mu_3$ can thus take any value between $\mu_1$
and $\mu_2$. Since the same holds for $\mu_4$, $R_{4pt}$ can be
negative (see Fig. 1 (c,d)). Using
\mbox{$R_{2pt}=(\mu_{1}-\mu_{2})/I$} and
\mbox{$R_{4pt}=(\mu_{3}-\mu_{4})/I$}, we find that $R_{4pt}$ takes
any value between \cite{Buttiker2}
%====================================================================
\begin{equation}
-R_{2pt}\leq R_{4pt}\leq R_{2pt}
\end{equation}
%====================================================================
Previous work on ballistic one-dimensional conductors fabricated
in semiconductors showed that $R_{4pt}$ can become slightly
negative \cite{Timp,Takagaki,Picciotto}. However, large negative
modulations of $R_{4pt}$ remain to be observed.

We have fabricated nanotube circuits with a new layout for
four-point measurements. First, $\sim1$ nm diameter SWNTs grown by
laser-ablation \cite{Thess} or chemical-vapor deposition
\cite{Hafner} are selected with an atomic force microscopy (AFM).
Voltage electrodes are then defined by positioning two MWNTs above
the SWNT using AFM manipulation. We choose such voltage electrodes
since the electric transmission between two nanotubes is known to
be low \cite{Fuhrer,Gao}. Then, Cr/Au electrodes are patterned for
electric connection with standard electron-beam lithography
techniques (see Fig. 1(a)).

\begin{figure}
\includegraphics{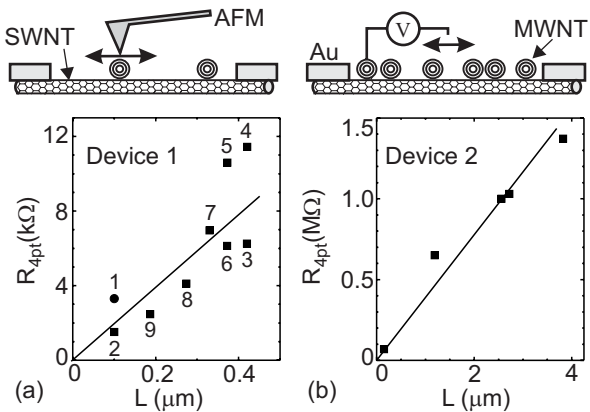}
 \caption{
Length dependence of $R_{4pt}$ at room temperature and $V_g=0$.
(a) SWNT contacted by 2 MWNTs. One MWNT is displaced back and
forth with an AFM tip. Points are numbered to describe the
measurement sequence. Point 1 has been acquired one week before in
the cryostat. (b) SWNT contacted by 6 MWNTs.
 }
\end{figure}

We now look to the length dependence of $R_{4pt}$ measured at
300K. Fig. 2(a,b) shows that $R_{4pt}$ linearly increases with the
length, in agreement with Ohm's law and eq. 1. Two types of
measurements have been carried. In Fig. 2(b), 6 MWNTs have been
placed on a long SWNT, enabling the resistance measurement of
multiple portions \cite{Remark2}. In Fig. 2(a), an AFM tip has
been used to change the separation between two MWNTs. Most points
have been recorded while decreasing $L$, so that the resistance
enhancement with the length is not due to a structural degradation
during the manipulation. However, the manipulation might well
stretch the tube, deposit or remove some molecules adsorbed on the
SWNT, or modify the pressure applied by the MWNT on the SWNT.
Those modifications may account for the scattered $R_{4pt}$ points
in Fig. 2(a).

The measurements above also show that $R_{4pt}$ tends to zero as
the length is reduced. This suggests no (or little) additional
contribution to the resistance from the MWNT contacts. Our device
layout thus allows for the first direct measurement of the
intrinsic resistance of a nanotube. The lowest resistance that we
obtained is \mbox{1.5 k$\Omega$} for a 95 nm long section (Fig.
2(a)). Such a low resistance is remarkable, since the resistance
of quasi one-dimensional conductors is expected to be dramatically
enhanced with the presence of disorder or phonons. This agrees
with previous two-point measurements \cite{Liang,Kong,Mann,Babic}
since $R_{2pt}$ was shown to approach the quantum resistance
$h/4e^2=6.5$ k$\Omega$, which suggests low intrinsic resistance.

Further insight into transport properties is obtained by
decreasing the temperature $T$. Figure 3 shows that $R_{4pt}$ does
not change for $T$ above $\sim80$ K, suggesting that the intrinsic
resistance is related to some static disorder and not to phonons.
At lower $T$, however, the inset of Fig. 3 shows that $G_{4pt}$
fluctuates with sweeping of the backgate voltage $V_g$.

\begin{figure}
\includegraphics{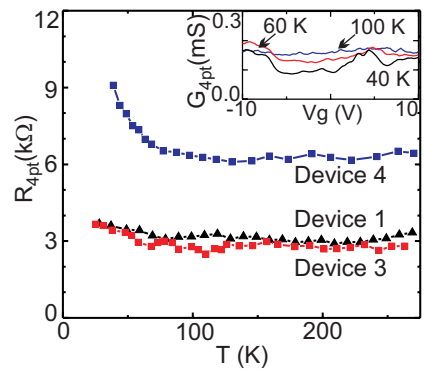}
 \caption{
Temperature dependence of $R_{4pt}$ measured at zero $V_g$. The
inset shows $G_{4pt}$ as a function of $V_g$ for Device 4.
 }
\end{figure}

We now discuss the possible origin of these fluctuations. They may
originate from low-transmission barriers created by the MWNTs or
some static disorder that form quantum dots along the tube. The
transport may then be dominated by Coulomb blockade (CB). However,
the conduction stays mostly constant with $T$ above
\mbox{$T^*\sim60$K}, the temperature at which fluctuations appear.
This is in opposition to CB theory \cite{Joyez} that predicts
\mbox{$G=G_0(1-\frac{1}{3}E_c/kT)$} for temperatures larger than
the charging energy $E_c$, expected to be close to $kT^*$.
Moreover, the best fit between this model and measurements above
$T^*$ gives $E_c/k\simeq 5$ K, which is much lower than $T^*$.
Another mechanism is thus needed to account for the fluctuations

The presence of disorder is expected to generate a complicated
interference pattern along the tube that should vary with the
Fermi level. This results in aperiodic conductance fluctuations
around the classical conductance when sweeping $V_g$
\cite{Beenakker}, which is consistent with our measurements. Such
fluctuations appear when effects of thermal averaging and phase
decoherence are weak enough so that the thermal length $L_T= \hbar
v_F/kT$ and the coherence length $L_\varphi$ are larger than the
elastic mean-free path. Interestingly, $L_T$ at $T^*=60$ K is
comparable to $l_e$ that is determined using Eq. 1 (see Table 1).
This suggests that thermal averaging is here at least as
detrimental as decoherence. Note, moreover, that the amplitude of
the fluctuations approaches $e^2/h$ at $\sim40$ K. This is
expected \cite{Beenakker} when the lower of $L_T$ and $L_\varphi$
is comparable to the separation between the voltage electrodes,
which is again consistent with measurements since $L_T=150$ nm and
$L=140$ nm. Overall, our measurements suggest that
quantum-interference effects start to modify the classical
resistance given by eq. 1 around a few tens of Kelvin.

\begin{table}
\includegraphics{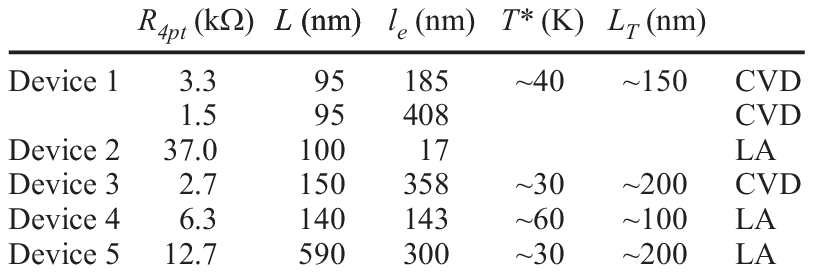}
 \caption{
Device characteristics. The first line of Device 1 corresponds to
properties measured in the cryostat and the second to that
measured one week later during AFM manipulation. Values of Device
2 are extracted from the linear fit in Fig. 2(b). All SWNTs are
metallic except Device 1 and 2. Device 1 is a small-gap
semiconductor with the current reduction occurring at $V_g>$2 V.
Device 2 is a large-gap semiconductor with the threshold voltage
at $~40$ V. LA = laser ablation. CVD = chemical vapor deposition.
 }
\end{table}

We turn now our attention to two-point measurements at the much
lower temperature of 1.4 K. Figure 4(a) shows a series of CB
peaks, which appear at the same gate voltages for measurements
between different pairs of electrodes. This indicates that Au and
MWNT electrodes probe the same nanotube quantum dot. The peak
occurrence is quite regular with a spacing of $V_g\simeq75$mV.
Coulomb diamonds measurements (not shown) give that the charging
energy $E_c\approx5$ meV. It has been shown that $E_c\approx
\frac{5 \textrm{meV}}{L[\mu m]}$ for similarly prepared samples
\cite{Nygard}, so that the dot length is $\sim1$ $\mu$m, which is
consistent with the 600 nm separation between Au electrodes. These
measurements suggest that MWNTs are sufficiently noninvasive to
not divide the SWNT in multiple quantum dots, in opposition to
nanofabricated electrodes \cite{Bezryadin}.

\begin{figure}
\includegraphics{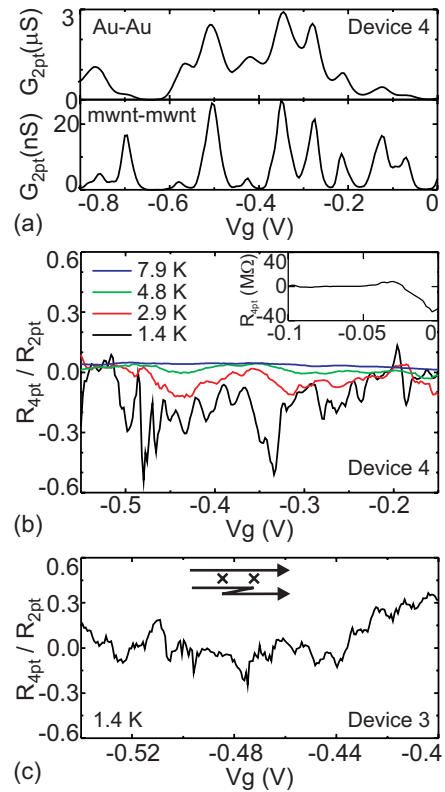}
 \caption{
Negative four-point resistance at low temperature. (a) Two-point
conductance as a function of $V_g$ at 1.4 K. $R_{2pt}$ measured
between Au electrodes is checked to be lower than $R_{2pt}$
between MWNT electrodes in this $V_g$ region. CB peaks measured
between different electrodes appear also at the same $V_g$s for
Device 1 and 3. (b,c) $R_{4pt}/R_{2pt}$ as a function of $V_g$.
The inset of (b) shows $R_{4pt}(V_g)$ at 1.4 K. Similar results
are also obtained for Device 1. The inset of (c) shows two
scattering centers that generate interference.}
\end{figure}

Having shown that the SWNT is a single quantum dot over its length
at 1.4~K, we now look at the four-point measurements. The inset of
Fig. 4(b) shows that $R_{4pt}$ becomes -29 M$\Omega$ near zero
$V_g$. However, $R_{4pt}$ significantly drops to 0 for $V_g$
between -0.05 and -0.1 V when the two-point conductance between Au
electrodes increases, in agreement with Eq. 3. Figures 4(b,c) show
that the modulations of $R_{4pt}/R_{2pt}$ are significant, with
absolute values that reach as high as 0.6.

We now discuss the origin of the negative $R_{4pt}$. It could
arise from voltage electrodes that differently couple to the left
and the right moving states inside the tube
\cite{Datta,Timp,Takagaki}. However, such a classical effect
should persist at higher temperature, which is not the case since
$R_{4pt}$ is only positive at $T\gtrsim 10~\textrm{K}$.

The resistance ratio $R_{4pt}/R_{2pt}$ is given by
\cite{Buttiker2}
%====================================================================
\begin{equation}
\frac{R_{4pt}}{R_{2pt}}=\frac{T_{31}T_{42}-T_{32}T_{41}}{(T_{31}+T_{32})(T_{41}+T_{42})}
\end{equation}
%====================================================================
Part of the fluctuations are expected to come from the Coulomb
blockade observed at 1.4 K that leads to oscillations in $T_{ij}$
transmissions. However, regular $V_g=75$mV CB oscillations cannot
alone account for the rapid $V_g\sim10$mV fluctuations of
$R_{4pt}/R_{2pt}$ in Fig. 4(b). We rather attribute those
fluctuations to quantum-interference terms \cite{Gopar,Buttiker3}
that are contained in $T_{ij}$ transmissions and that may arise
from the superposition of different electronic paths between $i$
and $j$ electrodes \cite{Remark3}. Indeed, the disorder along the
SWNT leads to different possibilities in the pathway between 2
electrodes (see the inset of Fig. 4(c)). It is also likely that
the sign change originates from those interferences. Variations of
$T_{ij}$ transmissions with $V_g$ are then uncorrelated, enabling
the sign reversal of the numerator in Eq. 4.

$R_{4pt}/R_{2pt}$ goes to zero when the temperature is increased
to $\sim5$K (Fig. 4(b)). This is expected when the nanotube is no
longer phase-coherent over its length, so that the
quantum-interference term of the transmissions vanishes. Another
possibility is that the quantum-interference term is washed out
due to thermal averaging, since $kT$ enhances the number of
electron paths between two electrodes that contribute to
transport. Interestingly, the 510 nm thermal length at 5K is
comparable to the 600 nm separation between Au electrodes of
Device 4. $L_T$ is $(\hbar v_Fl_e/kT)^{1/2}$ since the transport
is here diffusive. This points to the same finding observed at
higher temperature that the thermal length apparently is the
relevant parameter for quantum-interference phenomena in SWNTs.

In conclusion, we have shown that quantum-interference effects
dramatically modulate the four-point resistance of SWNTs, so that
$R_{4pt}$ can even become negative. This happens when the thermal
length is longer than the dimension of the system. The thermal
length is 20 nm long at room temperature; hence it is likely that
inclusion of these quantum-mechanical interference effects will
ultimately be required in the design of practical multi-terminal
intramolecular devices.

We thank C. Delalande for support, L. Forro for MWNTs, and R.
Smalley for laser-ablation SWNTs. LPA is CNRS-UMR8551 associated
to Paris 6 and 7. The research in Paris has been supported by ACN,
Sesame. YFC and MSF acknowledge support from the U.S. National
Science Foundation through grant DMR-0102950.

$^{*}$ corresponding author: bachtold@lpa.ens.fr


\begin{references}
\bibitem{Remark1}
Equation 1 is obtained from the Landauer formula $h/4e^2 (1-T)/T$
with the transmission $T=l_e/(L+l_e)$. Alternatively, eq. 1 can be
derived from the conductivity $\sigma =e^2D\rho$ with the
diffusion constant $D=v_Fl_e$ and the density of states
$\rho=4/(hv_F)$ for SWNTs.
\bibitem{Remark2}
We checked that $R_{4pt}$ at 300 K is independent of the contact
resistance at the current electrodes by using different pairs of
MWNT and Au electrodes for the current injection. $R_{2pt}$ varied
between 1.2 and 11 M$\Omega$.
\bibitem{Remark3}
Sweeping the magnetic field $B$ induces fluctuations of $R_{4pt}$
with the smallest fluctuations $\approx 6$~T for Device 4. The
corresponding Zeeman energy, which is 0.7~meV using the Land\'{e}
factor $g=2$, is consistent with the rapid $V_g\approx10$mV
fluctuations in Fig. 4(b) since the coupling efficiency measured
from Coulomb diamonds is $\alpha=1/15$.
\end{references}
\end{document}